\documentclass[aps,prl,superscriptaddress,showpacs]{revtex4}
\usepackage{amsmath}
\usepackage{graphicx,amssymb}
\usepackage{indentfirst,latexsym,bm}
\usepackage{threeparttable}
\begin{document}
\title{Decoherence of quantum gates based on Aharonov-Anandan phases in a multistep scheme}
\author{Xin Li}
\affiliation{State Key Laboratory of Surface Physics and  Department of Physics, Fudan University, Shanghai, 200433, China}
\affiliation{Faculty of Science, Kunming University of Science and Technology,
Kunming  650093, China}
\author{Yu Shi}
\email{yushi@fudan.edu.cn}
\affiliation{State Key Laboratory of Surface Physics and  Department of Physics, Fudan University, Shanghai, 200433, China}
\begin{abstract}
We study quantum decoherence of single-qubit and two-qubit Aharonov-Anandan (AA) geometric phase gates realized in  a multistep scheme. Each AA gate is also compared with the  dynamical phase gate performing the same unitary transformation within the same time period and  coupled with the  same  environment, which is modeled as harmonic oscillators.   It is found that the fidelities and the entanglement protection of the AA phase gates are enhanced by  the states being superpositions of different eigenstates of the environmental coupling, and the noncommutativity between the qubit interaction and the environmental coupling.
\end{abstract}
\maketitle

There exists the persistent motivation to realize quantum computation. Implementing  universal sets of quantum gates in terms of geometric phases  has been extensively studied, and both the adiabatic geometric phases, i.e. Berry phases,  and the nonadiabatic geometric phases, i.e.  AA phases, were considered  \cite{zanardi}. As the motivation for constructing geometric quantum gates is the presumed  insensitivity to noises, it is important to investigate the effects of both the classical fluctuations of the parameters and the quantum decoherence due to coupling with the environment. The robustness of the Berry phase in the adiabatic limit against the classical fluctuations of the parameters  has been demonstrated theoretically~\cite{chiara} and experimentally~\cite{leek,filipp,berger}. The experimentally observed decrease of dephasing with the increase of driving amplitude~\cite{leek} and the noise resilience in the adiabatic limit~\cite{leek,filipp,berger} is  in consistency with the suppression of nonadiabatic error by adiabaticity~\cite{childs}.    Using  superconducting qubits, a geometry dependent contribution to the dephasing of the Berry phase was observed~\cite{leek}, and it was demonstrated that  the Berry phase is less affected by the noise along the path than by that perpendicular to it~\cite{berger}. The behavior of the Berry phase under the quantum decoherence has also been studied theoretically, mostly with indications of some kinds of robustness under some conditions~\cite{berrydecoherence}.

To our knowledge, however, so far the behavior of AA phase gates under classical parametric noises or quantum decoherence has only been studied theoretically in a few papers~\cite{zhu,nazir,blais,li}. In a scheme of quantum gates based on one rotation of the qubit, in which the phase is a sum of the AA geometric phase and the dynamical phase,   it was clearly shown  that the fidelity  in presence of the parameter noise increases with the fraction of the geometric phase, and that  for the quantum decoherence, the gate fidelity was found to be determined only by the operation time~\cite{zhu}. A nonadiabatic geometric phase gate in  a  multistep scheme, which is designed to cancel the dynamical phases,  shows larger rates of entanglement loss and fidelity loss against quantum decoherence than a one-step dynamical phase gate, interpreted as due to longer operation time, but there is little sensitivity to noise when a large relative phase is generated between the qubit amplitudes~\cite{nazir}.      The AA phase gates  in another  multistep scheme were also found to be less robust  against the classical fluctuations of the parameters than the dynamical phase gates~\cite{blais}.  However, the results in Ref.~\cite{nazir,blais} were understood to be due to the fact that the geometric gates are completed in three rotations while the dynamical gates are completed in one step only~\cite{zhu}.

In this paper,   by modeling the environment as  harmonic oscillators, we study the quantum decoherence of single-qubit and two-qubit AA phase gates in a multistep scheme of the type considered in Ref.~\cite{blais}, where only the classical fluctuations of the parameters were considered. We also make the similar calculations for the equivalent dynamical gates coupled  to the same environment in the same way.  Here the two gates are referred to  as equivalent if  they  realize a same unitary transformation by using a same Hamiltonian for a same period of time.

Let us start with the following effective  Hamiltonian,
\begin{equation}\label{Eq2}
H=-\frac{1}{2} B_{z}\sigma_{z}- \frac{1}{2} B_{x}\sigma_{x},
\end{equation}
which is the standard one for the implementation  using a superconducting charge qubit, with the qubit basis states represented by two charge states of the Cooper pairs, $B_{z}$ proportional to the charging energy and and tunable  through the gate voltage,   $B_x$  proportional to the Josephson energy and tunable through the external magnetic flux~\cite{15}.  A same single-qubit unitary transformation
\begin{equation}
U_{single} = \left(
               \begin{array}{cc}
                 e^{-i\Phi} & 0 \\0 & e^{i\Phi} \\
               \end{array}
             \right) \label{u}
\end{equation}
can be realized either  in terms of  an AA phase  or  in terms of  a simple dynamical phase~\cite{blais}. Furthermore,  the two gates are completed in a same time period.

\begin{figure}
\scalebox{0.9}[0.9]{\includegraphics{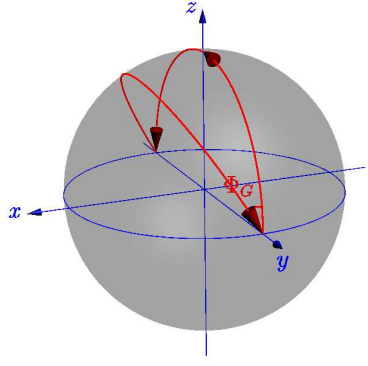}}
\caption{ \label{fig1} (Color online) The closed path traversed by the Bloch vector starting from  $|+\rangle_{z}$.  The cyclic evolution leads purely to an AA phase $-\Phi$. Similarly, starting from $|-\rangle_{z}$, the same unitary operation leads to purely an AA phase $\Phi$.}
\end{figure}

To realize the AA phase gate, one  performs a sequence of unitary transformations to implement $U_{single}$ as~\cite{blais}
\begin{equation}
U_{single}=U_{1}U_{2}U_{1}, \label{aasingle}
\end{equation}
where
$U_1 \equiv e^{-i\frac{\pi}{4}\sigma_{x}}$
is achieved by the evolution under  the Hamiltonian (\ref{Eq2}) with $B_{z}=0$ and $B_{x}=B_0$, where $B_0$ is some constant,  for a time period $ t_1 \equiv \pi\hbar/(2B_{0})$, $
U_2 \equiv e^{-i\frac{\pi}{2}(\sin\Phi\sigma_{z}-\cos\Phi\sigma_{x})}$
is achieved by  the evolution under the Hamiltonian  (\ref{Eq2})  with  $B_{z}=B_{0}\sin\Phi$ and  $B_{x}=B_{0}\cos\Phi$ for a time period  $2t_1$. This sequence of unitary transformations moves the Bloch vector of the state  along the geodesics on the Bloch sphere (Fig.~\ref{fig1}). In the end, the basis state $|\pm\rangle_{z}$ acquires a pure AA phase $\mp\Phi$, where $\Phi$ is equal to half the solid angle enclosed by the loop.

On the other hand,  if the qubit evolves under the Hamiltonian (\ref{Eq2}) with $B_{z}=\Phi\hbar/t_D$ and $B_{x}=0$ for a time period $t_D=4t_1$,  $U_{single}$ is realized in terms of dynamical phases only, hence starting from the initial state  $|\pm\rangle_{z}$, a simple unitary evolution $e^{-i\Phi\sigma_{z}}$ leads to a pure dynamical phase $\mp\Phi$~\cite{blais}.

Now we consider the decoherence originated from the coupling with  a bath of harmonic oscillators. For a superconducting charge qubit, the voltage fluctuation coupled to the $\sigma_{z}$  is dominant \cite{15,16}. Hence the total Hamiltonian of the composite system can be written as
\begin{equation}
H_{T}=H+H_{I}+H_{E}, \label{Eq3}\end{equation}
where $H$ is given in (\ref{Eq2}),
$H_{I}=\sigma_{z}\sum\limits_{k}g_{k}(a_{k}^{\dag}+a_{k})$, $H_{E}=\sum\limits_{k}\omega_{k}a_{k}^{\dag}a_{k}$. One obtains the well-established Bloch-Redfield equation for the density matrix of the qubit \cite{17},  $\dot{\rho}=-\frac{i}{\hbar}[H,\rho]-\frac{1}{\hbar^{2}}\int_{0}^{\infty}ds
Tr_{E}[H_{I},[H'_{I}(-s),\rho\otimes\rho_{E}],$
where $H'_{I}(-s)=e^{-i(H+H_{E})s/\hbar}H_{I}e^{i(H+H_{E})s/\hbar}$, $Tr_{E}$ represents the partial trace over the environment, which is assumed to be initially in thermal equilibrium at temperature T, i.e., $\rho_{E}=\prod\limits_{k}(1-e^{-\omega_{k}/kT})e^{-\omega_{k}a_{k}^{\dag}a_{k}/kT}$. We obtain \begin{equation}\rho(t)=\exp(Mt/\hbar)\rho(0),\end{equation} where $M$ is a superoperator.

For the AA phase gate (\ref{aasingle}),
 \begin{widetext}
 \begin{equation}\label{Eq9}
M=\left(
       \begin{smallmatrix}
          -\sin^{2}\theta \mu_{+}(B) & \frac{1}{2}\sin(2\theta)\mu_{+}(0) & \frac{1}{2}\sin(2\theta)\mu_{+}(0) & \sin^{2}\theta \mu_{+}(-B) \\
           \frac{1}{2}\sin(2\theta)(\mu_{-}(0)+2\Gamma^{*}(B)) & -2iB-2\cos^{2}\theta\mu_{+}(0)-\sin^{2}\theta\xi_{+}^{*}(B)  & \sin^{2}\theta\xi_{+}(B)& \frac{1}{2}\sin(2\theta)(\mu_{-}(0)-2\Gamma(-B)) \\
           \frac{1}{2}\sin(2\theta)(-\mu_{-}(0)+2\Gamma(B)) & \sin^{2}\theta\xi_{+}^{*}(B)& 2iB-2\cos^{2}\theta\mu_{+}(0)-\sin^{2}\theta\xi_{+}(B) & -\frac{1}{2}\sin(2\theta)(\mu_{-}(0)+2\Gamma^{*}(-B)) \\
           \sin^{2}\theta \mu_{+}(B) & -\frac{1}{2}\sin(2\theta)\mu_{+}(0) & -\frac{1}{2}\sin(2\theta)\mu_{+}(0) & -\sin^{2}\theta \mu_{+}(-B) \\
       \end{smallmatrix}
     \right),
 \end{equation}
\end{widetext}
where $ B=\sqrt{B_{z}^{2}+B_{x}^{2}}$,  $\tan\theta \equiv B_{x}/B_{z}$,
 $\mu_{\pm}(B) \equiv \Gamma(B)\pm\Gamma^{*}(B)$, $\xi_{\pm}(B) \equiv \Gamma(-B)\pm\Gamma^{*}(B)$, with
$\Gamma(B)=\int_{0}^{\infty}d\tau\int_{0}^{\Omega}d\omega J_{o}(\omega)e^{2iB\tau}[\coth\frac{\omega}{2kT}\cos(\omega\tau)-i\sin(\omega\tau)]
=\frac{\pi}{2}J_{o}(2B)(\coth(\frac{B}{kT})+1)
-i\mathbf{P}\int_{0}^{\Omega}d\omega\frac{J_{o}(\omega)}{\omega^{2}-4B^{2}}(2B\coth(\frac{\omega}{2kT})-\omega).$
 Here $\mathbf{P}$ denotes the principal value, the noise spectral density is assumed to be Ohmic, i.e.
$
J_{o}(\omega)=\lambda\frac{\omega}{1+\frac{\omega^{2}}{\Omega^{2}}},
$
where $\lambda$ is the dimensionless coupling strength, $\Omega$ is the cutoff frequency.

\begin{figure}
\scalebox{0.6}[0.6]{\includegraphics{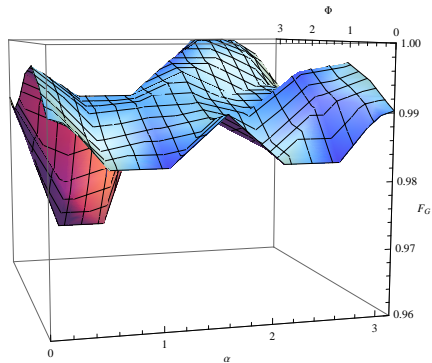}}
\caption{\label{fig2} (Color online) The fidelity of the geometric phase gates as a function of the parameters $(\alpha,\Phi)$ at $T=0$.  The parameter values are $B_0/\hbar=10GHz$, $\Omega=50B_0$, $\lambda=10^{-3}$,  obtained from the  usual experimental values for superconducting qubits.}
\end{figure}

For an arbitrary initial state    $\cos\alpha|+\rangle_{z}+\sin\alpha|-\rangle_{z}$, the action of  $U_{1}U_{2}U_{1}$ leads to the final state $e^{-i\Phi}\cos\alpha|+_{z}\rangle+ e^{i\Phi}\sin\alpha|-_{z}\rangle$, with a pure AA phase $\Phi$.   In order to see the  decoherence effect on the pure AA phase gate, we plot the fidelity of the final state  as a function of parameters $(\alpha,\Phi)$ in Fig.~\ref{fig2}, which indicates the decoherence even at zero temperature. We only need to consider $0\leq \alpha\leq \pi$, since for $\pi+\alpha$, the initial density matrix $\rho(0)$ is the same as for $\alpha$.

\begin{figure}
\scalebox{0.75}[0.75]{\includegraphics{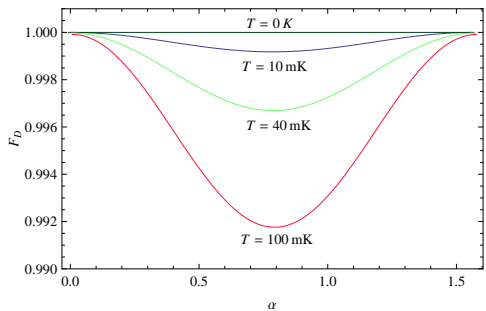}}
\caption{\label{fig3} (Color online) The fidelity of the dynamical phase gates as a function of $\alpha$ at various temperatures.}
\end{figure}

\begin{figure}
\scalebox{0.9}[0.9]{\includegraphics{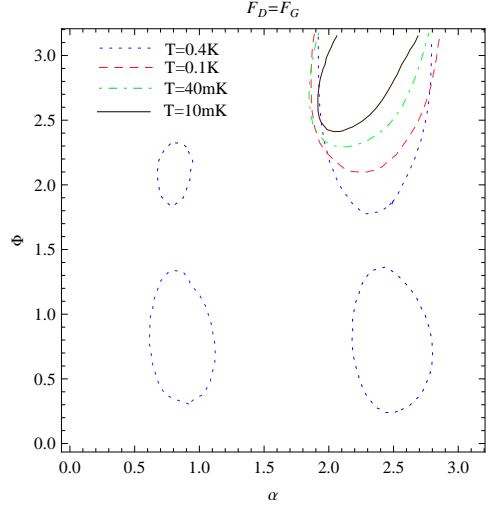}}
\caption{\label{fig4} (Color online) The contours for the equality of  the fidelities of  the AA geometric phase gate and the dynamical phase gate realizing a same single-qubit unitary operation within a same time period. Only  inside  the closed curves are the AA phase gates  less affected by decoherence than  the dynamical ones.}
\end{figure}

For the equivalent dynamical phase gate,
\begin{equation}\label{Eq12}
M=\left(
       \begin{matrix}
         0 & 0 & 0 & 0 \\
          0 & -2iB-2\mu_{+}(0) & 0 & 0 \\
           0 & 0 & 2iB-2\mu_{+}(0) & 0 \\
            0 & 0 & 0 & 0 \\
           \end{matrix}
     \right),
      \end{equation}
where $\mu_{+}(0)=2\pi kT\lambda$. Consequently the fidelity of the dynamical phase gate can be easily obtained as
 $
 F_{D}=\cos^{4}\alpha+\sin^{4}\alpha+2e^{-4\pi kT\lambda t_{d}/\hbar}\cos^{2}\alpha\sin^{2}\alpha,$ which is  shown in Fig.~\ref{fig3}.   $F_{D}$ reaches the minimum at $\alpha=\pi/4$ for the finite temperature. Moreover,   $F_D$ is independent of $\Phi$, as $M$ is diagonal,  because of the commutativity between the qubit Hamiltonian and the coupling with environment.
The dynamical phase gate is  immune to  decoherence at zero temperature, as  the Hamiltonian of the qubit now commutes with the coupling with the bath and thus there is no energy dissipation, while there is no dephasing because of the zero temperature. In  Fig.~\ref{fig4}, we draw the contours of  $F_{D}=F_{G}$ in the plane of $\alpha$ and $\Phi$. $F_{D} < F_{G}$ inside each contour curve, while $F_{D} > F_{G}$ in the regimes outside these  contour curves.

\begin{figure}
\scalebox{0.95}[0.95]{\includegraphics{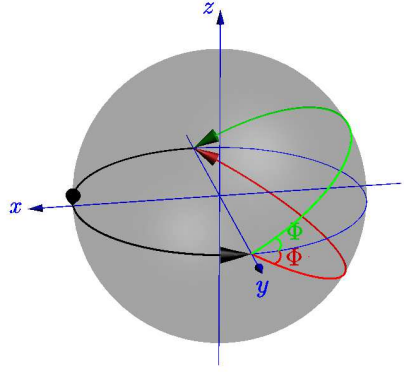}}
\caption{\label{fig5} (Color online) The paths on the Bloch sphere of qubit-1 starting with $|+\rangle^{(1)}_{x}$. When  qubit-2 is in the state $|+\rangle^{(2)}_{x}$ ($|-\rangle^{(2)}_{x}$), the state of qubit 1 completes a cyclic evolution along the half circle with $z=0$ and $x >0$ plus the half circle with $z <0$  ($z >0$ ),  and in the end acquires  a AA phase $\Phi=\pi-\Phi$ ($\Phi=\pi+\Phi$).}
\end{figure}

The above schemes for single-qubit gates can be generalized to two-qubit gates based on the effective Hamiltonian
\begin{equation}\label{Eq14}
H_{12}=\sum\limits_{i=1,2}[-\frac{1}{2}B^{(i)}_{z}\sigma^{(i)}_{z}
-\frac{1}{2}
B^{(i)}_{x}\sigma^{(i)}_{x}]
-J\sigma^{(1)}_{x}\sigma^{(2)}_{x}.
\end{equation}
This Hamiltonian  has been experimentally realized in terms of superconducting charge qubits~\cite{13}, where $B^{(i)}_{z}$, $B^{(i)}_{x}$ and $J$ can be tuned. Using this Hamiltonian, we can realize
\begin{equation}
U_{two} = e^{i(\pi-\Phi)}|++\rangle_x {_x}\langle ++| + e^{i(\pi+\Phi)}|+-\rangle_x {_x}\langle +-| +
e^{-i(\pi-\Phi)}|-+\rangle_x {_x}\langle - +| +
e^{-i(\pi+\Phi)}|--\rangle_x {_x}\langle - -|
\end{equation}
in terms of  an AA phase  or  in terms of  a simple dynamical phase.
By applying this conditional geometric phase operation together with certain single-qubit operations, the controlled-NOT gate, with respect to the basis  $\{|\pm\rangle^{(1)}_{x}|\pm\rangle^{(2)}_{x}\}$, can be realized as $C_{NOT}=e^{i\frac{\pi}{2\sqrt{2}}(\sigma_{z}^{(2)}-\sigma_{x}^{(2)})}e^{i\frac{\pi}{4}
\sigma_{x}^{(1)}}e^{i\frac{\pi}{4}\sigma_{x}^{(2)}}U_{two}(\frac{\pi}{4})e^{i\frac{\pi}{2}\sigma_{x}^{(1)}}
e^{i\frac{\pi}{2\sqrt{2}}(\sigma_{z}^{(2)}-\sigma_{x}^{(2)})}$, up to a global phase factor.

By tuning  $H_{12}$, one can realize a two-qubit unitary operation $U_{two}$ as  a conditional geometric phase gate  through a unitary sequence,
$$U_{two}= U^{(1)}_{1}U^{(12)}_{2}U^{(1)}_{1},$$
where $U^{(1)}_1 \equiv e^{-i\frac{\pi}{4}\sigma^{(1)}_{z}}$ is   a single-qubit operation  on qubit-1, $U^{(12)}_{2} \equiv e^{-i\frac{\pi}{2} (\cos\Phi\sigma^{(1)}_{z}-\sin\Phi\sigma^{(1)}_{x}\sigma^{(2)}_{x})}$ is a two-qubit operation   and  can be regarded as a unitary operation on qubit-1  conditional on the state of qubit-2.  $U^{(1)}_1$ is realized by setting $B^{(2)}_z=B^{(2)}_x=B^{(1)}_x=J=0$ while $B^{(1)}_{z}=2J_{m}$, where $J_m$ is a certain constant value, for a time period $t_{m1}=\frac{\pi\hbar}{4J_{m}}$.  $U^{(12)}_{2}$ is  realized by setting $B^{(2)}_z=B^{(2)}_x=B^{(1)}_x=0$ while  $B^{(1)}_{z}=2J_{m}\cos\Phi$, $J=J_{m}\sin\Phi$ for a time period $2t_{m1}$. The total time duration is $4t_{m1}$.   After undergoing the unitary sequence as depicted in Fig.~\ref{fig5}, $U_{two}$ is achieved in terms of AA phases.

The equivalent conditional dynamical phase gate realizing $U_{two}$ is simply $$U_{two}=e^{i(\pi-\Phi)\sigma^{(1)}_{x}\sigma^{(2)}_{x}},$$ which   can be  implemented simply by setting $J=\frac{(\pi-\Phi)J_{m}}{\pi}$ for a time period  $4t_{m1}$.

Now let us address the decoherence of the two-qubit gates.
According to the different correlation lengths of the noise sources, the dissipative environment is conveniently simulated as either a common bath of harmonic oscillators, which affects the two charge qubits in a same manner,  or  two independent   baths of harmonic oscillators, which affect the two charge qubits independently. The   interaction plus the bath Hamiltonian  can be written  as \cite{20}
\begin{equation}\label{Eq16}
 H_{I}+H_{E}= \left\{\begin{array}{l}
   (\sigma^{(1)}_{z}+\sigma^{(2)}_{z})
   \sum\limits_{k}[g_{k}(a_{k}^{\dag}+a_{k})+\omega_{k}a_{k}^{\dag}a_{k}],  {\rm  \mbox{ } one \mbox{ } common \mbox{ } bath, } \\
 \sum\limits_{i=1,2}\sum\limits_{k}[\sigma^{(i)}_{z}g^{(i)}_{k}
 (a^{(i)\dag}_{k}+a^{(i)}_{k})+\omega^{(i)}_{k}a_{k}^{(i)\dag}a_{k}^{(i)}], {\rm  \mbox{ }
two \mbox{ } indepent \mbox{ } baths.}  \end{array}\right.\\
 \end{equation}

\begin{figure}
\scalebox{1}[1]{\includegraphics{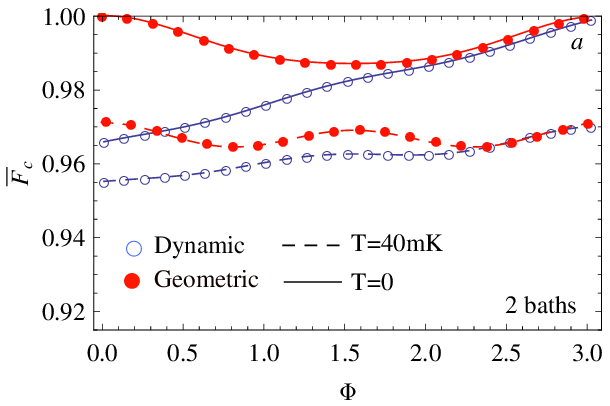}}
\scalebox{1}[1]{\includegraphics{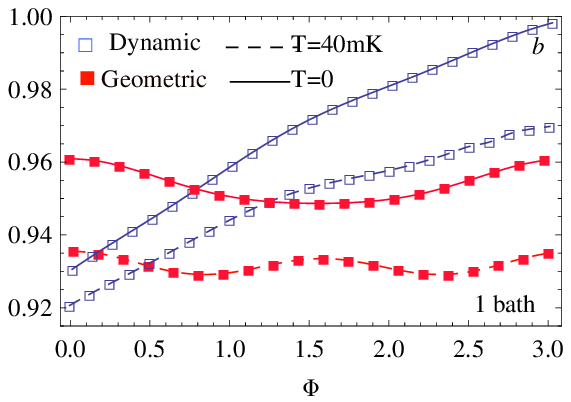}}
\caption{\label{fig6} (Color online) The average fidelities of the conditional geometric  phase gate and the conditional dynamical phase gate,  as functions of parameter $\Phi$. For the two-bath  case, $J^{(1)}_{o}(\omega)=J^{(2)}_{o}(\omega)$ is assumed. We always set $\lambda=10^{-3}$, $J_{m}/\hbar=5GHz$.}
\end{figure}

Considering various possible initial states, we calculate the average fidelity of each   two-qubit gate
$ \overline{F}_{two}=\overline{\langle\Psi_{in}|U^{\dag}_{two}
 \rho_{out}U_{two}|\Psi_{in}\rangle},$
where   $\rho_{out}$ is density matrix after applying the noisy gate to $|\Psi_{in}\rangle$, calculated by using the  Bloch-Redfield formalism above,  the overline indicates an average over a set of $10^{3}$ input states uniformly distributed over the parameters characterizing a two-qubit state.  From Fig.~\ref{fig6}, we note that the average fidelity is a monotonic function of the parameter $\Phi$ for the dynamical conditional gate, but is nonmonotonic for the equivalent AA conditional phase gate. This seems to be a reflection of the global character the geometric phase in contrast with the  dynamical phase, which is monotonic with the time. Moreover, the one-bath and two-bath cases show different features. In the two-bath case,  the geometric conditional phase clearly has larger average fidelity for most of the values of $\Phi$.

\begin{widetext}
\begin{figure*}
\scalebox{1.19}[1.19]{\includegraphics{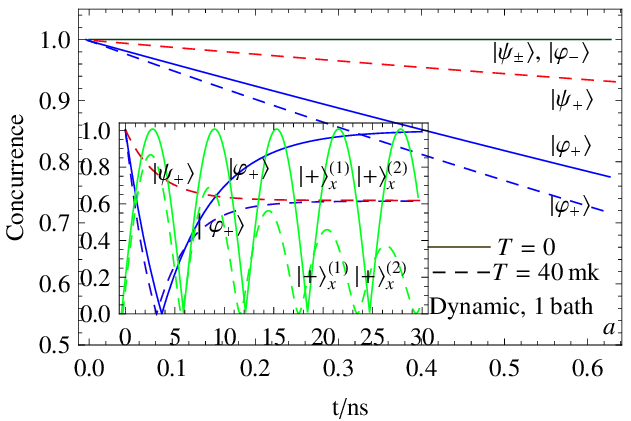}}
\scalebox{1.15}[1.15]{\includegraphics{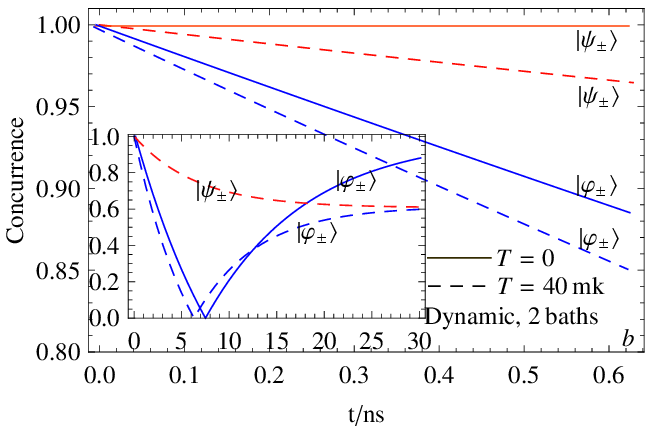}}
\scalebox{1.22}[1.22]{\includegraphics{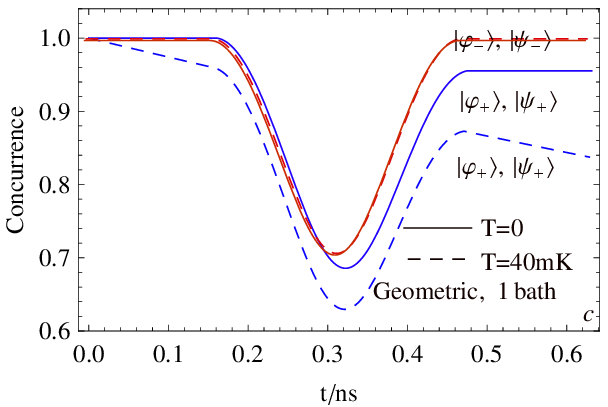}}
\scalebox{1.31}[1.31]{\includegraphics{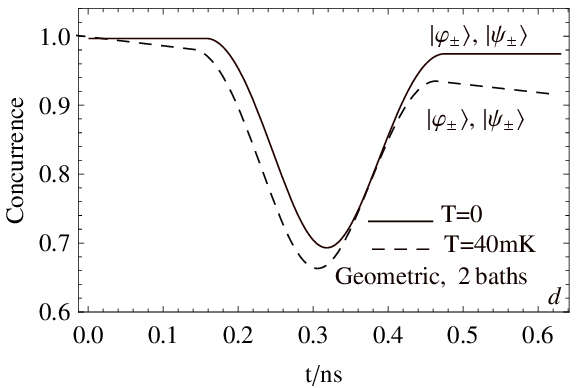}}
\caption{\label{fig7} (Color online) The evolution of the entanglement when  the Bell states $|\psi_{\pm}\rangle$ and $|\varphi_{\pm}\rangle$ are input to the equivalent  conditional geometric phase gates and the dynamical phase gates coupled with the same noise environment, respectively.  We  set $\Phi=\pi/4$, $\lambda=10^{-3}$, $J_{m}/\hbar=5GHz$. For two distinct baths, we assume $J_{o}^{(1)}(\omega)=J_{o}^{(2)}(\omega)$. The insets of a and b show the long-time behavior of the dynamical phase gates for the different initial states.}
\end{figure*}
\end{widetext}

We have also studied how the entanglement between the two qubits behaves under the coupling with the environment. Here the entanglement is quantified as the  concurrence  \cite{21}.   Fig.~\ref{fig7}  shows the evolution of the concurrence   under the action of the  noisy quantum gates, starting with each of the Bell states, $|\varphi_{\pm}\rangle=\frac{1}{\sqrt{2}}(|+\rangle^{(1)}_{x}|-\rangle^{(2)}_{x}
\pm|-\rangle^{(1)}_{x}|+\rangle^{(2)}_{x})$ and $|\psi_{\pm}\rangle=\frac{1}{\sqrt{2}}(|+\rangle^{(1)}_{x}|+\rangle^{(2)}_{x}
\pm|-\rangle^{(1)}_{x}|-\rangle^{(2)}_{x})$. Note that for the evolution starting from each Bell state under the action of $U_{two}$ in absence of the coupling with an environment, the entanglement always remains equal to  $1$.

Fig~\ref{fig7}a shows the result for the  the  dynamical gate in the one-bath case. The oscillating curves in the inset  indicate that  entanglement is generated even though the initial state is the non-entangled state $|+\rangle^{(1)}_{x}|+\rangle^{(2)}_{x}$. The concurrence oscillates with constant amplitude at zero temperature while with damped amplitude at finite temperatures. The decreased or disappeared entanglement  may revive on a long time scale.  For instance, for  $\rho(0)=|\varphi_{+}\rangle\langle\varphi_{+}|$, we  have
$
\rho(t)=P(t)|\psi_{+}\rangle\langle\psi_{+}|+
Q(t)|\varphi_{+}\rangle\langle\varphi_{+}|,$
where
$P(t) = \frac{\mu_{+}(J)(1-e^{-4t(\mu_{+}(J)+\mu_{+}(-J))/\hbar})}{\mu_{+}(J)+\mu_{+}(-J)}, $
 $Q(t) = \frac{\mu_{+}(-J)+\mu_{+}(J)e^{-4t(\mu_{+}(J)+\mu_{+}(-J))/\hbar}}{\mu_{+}(J)+\mu_{+}(-J)}.$
At zero temperature, $P(t)=1-e^{-8\pi tJ_{o}(2J)/\hbar}$ and $Q(t)=e^{-8\pi tJ_{o}(2J)/\hbar}$, indicating  the entanglement revival in long-time behavior, with $\rho(\infty)=|\psi_{+}\rangle\langle\psi_{+}|$.

For the conditional AA phase gate, as shown in Fig.~\ref{fig7}c and Fig.~\ref{fig7}d, the direct coupling of the two qubits in the operation  leads to the entanglement variation with a trough during the action of $U^{(12)}_2$,  while with small variations during the two actions of the single-qubit operation $U^{(1)}_1$. Overall speaking,  in sufficiently long time scale, the conditional AA phase gates are  advantageous over the equivalent dynamical phase gate in protection of quantum entanglement. It appears that   the direction of the noise source is quite relevant, as replacing the charge noise coupled to $\sigma_{z}$ with the flux noise coupled to $\sigma_{x}$ leads to a different situation.

To summarize,  we have studied the quantum decoherence  of  AA  phase gates in a multistep scheme, as well as that of the equivalent  dynamical phase gates realizing the same unitary  transformations in  same periods of time and coupled with the same environments.  Using the  Markovian Bloch-Redfield formalism,  we have calculated the fidelities of the single-qubit and two-qubit gates, as well as the time-dependent entanglement in two-qubit gates.

In our calculations, it appears that the robustness of a AA phase gate in this scheme is enhanced if the state is a superposition of different eigenstates of the environmental coupling. This feature is consistent with the previous result that there is little sensitivity to noise when a large relative phase is generated between the qubit amplitudes~\cite{nazir}.  This can be understood as the cancelation of the decoherence effect  because of the superposition or coherence feature of the state.    For example, when $\alpha$ is near $0$, $\pi/2$ and $\pi$, the state  $\cos\alpha|+\rangle_{z}+\sin\alpha|-\rangle_{z}$ is near an eigenstate of $\sigma_z$, which is proportional to the qubit-environment coupling, rather than a superposition of different eigenstates of $\sigma_z$,  the fidelity of the AA phase gate is less than that of the dynamical phase gate  in this regime (Fig.~\ref{fig4}). Another example is the following. The average fidelity of the two-qubit AA phase gate  in the case of coupling with  two independent baths  is larger  than the case of coupling with a same bath   can be understood to be related to stronger coherence in the two-bath case.  Yet another example is the following.  For the two-qubit AA phase gate with the interaction $\sigma_x\sigma_x$,    the entanglement protection in the case in which the environmental coupling is proportional to $\sigma_z$ is stronger than in the case in which the environmental coupling is proportional to $\sigma_x$, as can be explained  in terms of the noncommutativity between the qubit interaction and the environmental coupling, which leads to  the two-qubit state being a superposition of different eigenstates of  $\sigma_z$. Likewise,  in  the previous study using $\sigma_z\sigma_z$ interaction~\cite{nazir},  the loss of  fidelity and the decrease of entanglement are both larger in the case of  environment coupling  proportional to  $\sigma_z$ than in the case  of environmental coupling proportional to $\sigma_x$. This  can  also be  interpreted as that the robustness is enhanced by the  noncommutativity between the qubit interaction and the environmental coupling in the case of environmental coupling proportional to $\sigma_x$,  which leads to the  two-qubit state being a superposition of different eigenstates of  $\sigma_x$.

\acknowledgments

This work was supported by the National Science Foundation of China (Grant No. 11074048) and the Ministry of Science and Technology of China (Grant No. 2009CB929204).

\end{document}